\documentclass[epj,final]{svjour}[16-11-2012]
\usepackage{amssymb}
\usepackage{amsmath}
\usepackage{graphicx}
\usepackage{hyperref}
\usepackage{epsfig}
\usepackage{color}
\usepackage{dcolumn}
\usepackage{bm}
\usepackage{epstopdf}

\begin{document}

\title{Occurrence conditions for two-dimensional Borromean systems}
\author{A.G. Volosniev, D.V. Fedorov, A.S. Jensen, and N.T. Zinner}
  \institute{Department of Physics and Astronomy - Aarhus University, Ny Munkegade, bygn. 1520, DK-8000 \AA rhus C, Denmark}
\date{\today}

\abstract{
We search for Borromean three-body systems of identical bosons
in two dimensional geometry, i.e.
we search for bound three-boson system without bound two-body subsystems. 
Unlike three spatial dimensions, in two-dimensional geometry  
the two- and three-body thresholds 
often coincide ruling out Borromean systems.
We show that Borromean states can only appear for potentials 
with substantial attractive and repulsive parts. 
Borromean states are most easily 
found when a barrier is present outside an attractive pocket.
Extensive numerical search did not reveal Borromean states for potentials
without an outside barrier. 
We outline possible experimental setups 
to observe Borromean systems in two spatial dimensions.   
}
\authorrunning{A.~G. Volosniev {\it et al.}}
\titlerunning{Occurrence conditions for two-dimensional Borromean systems}
\maketitle

\section{Introduction}

In quantum mechanics two particles in three spatial dimensions (3D) can 
attract each other without forming a bound state. However adding a third
particle can make a three-body system bound. Such bound
three-body structures, where each pair of particles is unbound, are
called Borromean states.  They turn out to be rather abundant
in subatomic physics 
where many examples are found and studied, see e.g. \cite{zhukov1993,riisager2012}.
Those examples raised the question about properties of the potentials
that determine the possibility for Borromean binding. 
One can trace this discussion for three-dimensional geometry 
in numerous papers, e.g. \cite{richard1994,richard2000}. 
It turns out that in 3D finite range potentials  
of the form $gV(r)$ most likely have a region of the parameter $g$
where three, but not two particles are bound. 

In two spatial dimensions (2D), however, this question is not well-established. 
The result has to be different, 
because two-body binding is easier to
achieve in 2D than in 3D.  
This is shown already for two particles with $gV<0$.  Such 
potentials support two- and three-body bound states even when 
$g\rightarrow 0$ \cite{landau1977}.  
The thresholds for
binding of two and three-body systems are identical and
Borromean structures cannot exist.
Moreover, numerical investigations in momentum space
\cite{tjon1980,frederico1988} and coordinate space 
\cite{cabral1979,lim1980,nielsen1999,hammer2004,blume2005,hammer2011}
revealed that in 2D some other classes of potentials
have no Borromean states. 

Previous investigations also show, that it is not entirely impossible 
to have Borromean systems in 2D. However, to the best of our knowledge, only one 
example of an appropriate two-body potential can be found in the literature \cite{nielsen1999}. 
The purpose of this report is to identify conditions for the two-body
potentials that can produce Borromean systems of three identical bosons in two
dimensions.    
We believe that using advanced experimental techniques it is possible
to create setups where these conditions are satisfied. 
This introduces structures with new few- and many- body
properties.
This investigation also puts limitations for a zero-range 
formalism that is used widely to describe 
few-body dynamics in 2D \cite{kart2006,bel2011,bel2012,bel2013}. This formalism
is a powerful tool which has been proven to give 
correct results for weakly-bound systems in 2D that have identical three and two body 
thresholds. 
However it can not describe Borromean states where
finite-range techniques have to be used.

We believe that this investigation is timely and important. First, because the
experimental techniques in cold atomic gases are rapidly improving in 3D
and are currently being adjusted to 2D. Second, a Borromean system has
three particles which strongly suggests that
traditional many-body correlations built on two-body properties now
may turn out to be completely different if three particles
are considered as building blocks. Our results provide guidence to 
the tuning of the potentials in order to achieve new many-body structures.

In this report we briefly discuss
conditions for two-body binding  
in section II.   
We search for characteristic properties, e.g. shape of the potential
$V$ and strength $g$, which allow binding
of three but not of two particles in section III. 
We outline experimental setups to observe Borromean states in section IV.
Finally, in section V we briefly
summarize and conclude.

\section {Two-body problem in two dimensions}
According to the definition Borromean states, if they exist, appear
for two-body potentials at the edge of two-body binding.
 An additional particle
then provides the glue to form a three-body bound
state.
We therefore first must establish the criteria for a 
two-body potential  to support a bound
state \cite{first_footnote}.
This is known in the limit of weak potentials, $g\rightarrow 0$, where the
net volume, $g\int V r \mathrm{d}r$, simply has to be non-positive.  
However, the existence of
bound states for stronger potentials is not solely determined by the
net volume.  Thus we reformulate the two-body problem with the
aim of extracting suitable criteria for binding.  The qualitative
behavior is supplemented by a full quantitative analysis for simple
potentials.

\subsection {General properties}

The radial Schr\"{o}dinger equation for two particles in 2D is
\begin{equation}
  \frac{\hbar^2}{2\mu} \bigg(-\frac{1}{r}\frac{\partial}{\partial r}r\frac{\partial}{\partial r}+\
\frac{M^2}{r^2}\bigg)\phi_M  = (E_2-gV) \phi_M\; ,
 \label{shr-rad-eq}
\end{equation}
where $r$ is the relative coordinate, $E_2$ the two-body energy, $gV(r)$ the
potential, $\mu$ the reduced mass, and $M=0,1,2,...$ is the angular
quantum number, $\phi_M$ the wave function of the relative motion.
For spherical potentials $M$ is conserved and $M=0$
therefore characterizes the ground state for identical bosons.  We
shall in the following only consider $M=0$ and omit any related index.

We demand the wave function, $\phi(k,r)$, to be regular in zero, 
through  $\phi(k,r=0)=1 $, where $k$ is the wave number given by
$E_2=\hbar^2 k^2/(2\mu)$.  The condition for a bound state can then be
written as \cite{vol10,vol11}
\begin{eqnarray}  \label{jost-func}
 1+\frac{2g\mu}{\hbar^2}\frac{i\pi}{2}\int_0^\infty \mathrm{d}r r\phi(k,r)V(r)H_0^{(1)}(k r) = 0\;,
\end{eqnarray}
where $H_0^{(1)}$ is the first Hankel function of order zero. The
solutions correspond to bound states with $k=i|k|$.  We focus on the
weak binding limit where $k$ approaches zero and eq.~\eqref{jost-func}
reduces to
\begin{eqnarray}
1+\frac{2g\mu}{\hbar^2}\int_0^\infty \mathrm{d}r r\phi(0,r)V(r)
 \big(i\frac{\pi}{2} - \ln(\frac{k r}{2}e^\gamma)\big) =0\;,
 \label{jost-small}
\end{eqnarray}
where $\gamma$ is Euler's constant and the regular zero energy solution to
eq.~\eqref{shr-rad-eq} is
\begin{equation}
\phi(0,r)=1+\frac{2g\mu}{\hbar^2}\int_0^r \mathrm{d}s
s\ln{(r/s)}V(s)\phi(0,s) \;.
 \label{funct-small}
\end{equation}
Then eq.~\eqref{jost-small} can only be fulfilled for small $k$ when
\begin{equation}
g\int_0^\infty \mathrm{d}r r\phi(0,r)V(r)=0 \;.
 \label{cond-gen}
\end{equation}
Eqs.~\eqref{jost-small} and \eqref{funct-small}
define appearence of the bound state in 2D.
This result might also be obtained using the Jost function formalism, see 
\cite{gibson1986}.

From eq.~\eqref{shr-rad-eq} we get immediately 
\begin{equation}
g\int_0^R \mathrm{d}r r\phi(0,r)V(r) = \frac{\hbar^2}{2\mu}
\bigg(r\frac{\partial}{\partial r}\phi(0,r)\bigg)\bigg|_{r=R} = 0 \;,
\label{shr-eq-0}
\end{equation}
where the vanishing result is obtained for finite-range potentials where
$V(r>R)=0$.  Thus the slope of the zero-energy wave function is zero
at $r=R$.

It is obvious that for purely attractive potentials, $gV<0$,
the condition from eq.~\eqref{cond-gen} can be achieved 
only for $g=0$. This is a consequence of the  
fact that purely attractive potentials
always provide at least one two-body bound state in 2D \cite{landau1977}.
In this case the thresholds for binding of 
two and three-body systems has to coincide  
hence Borromean states are ruled out \cite{nielsen1999}.

It gives us the first necessary condition for Borromean binding: 
the potential, $gV$,  has to contain both positive and negative parts.  We shall
proceed by defining shapes $V_{\pm}$ and dimensionless strengths $\lambda_{\pm}>0$ 
of the positive and negative parts. 
The total potential is then
given by $gV(r)=\lambda_{+}V_{+} + \lambda_{-} V_{-}$. 
Binding is subsequently achieved by sufficient increase of $\lambda_{-}$, and
binding is reduced by increase of $\lambda_{+}$.  We define the
critical repulsive interaction, $\lambda_+^{(cr)}(\lambda_-)$, through
eq.~\eqref{cond-gen} by
\begin{equation}
\int_0^\infty\mathrm{d}r r \phi(0,r)(\lambda_+^{cr}V_+ + \lambda_-V_-)=0 \;.
 \label{condition}
\end{equation}
This implies that a Borromean system can only appear when $\lambda_+$
is larger than $\lambda_+^{cr}$, since two particles are unbound
in this regime.  Hence a Borromean system for given $\lambda_{-}$ might be
found in an interval where $\lambda_+$ is larger than
$\lambda_+^{cr}$, although perhaps an interval of very limited
extension.  The solution of eq.~\eqref{funct-small} and the definition
in eq.~\eqref{condition} then provide crucial information about the
most likely region for occurrence of Borromean systems.

Let us use eq.~\eqref{condition} for potentials, where 
knowledge of the wave function is not needed. First
we show the well established result that any weak potential
with negative or zero net volume has at least one bound state
\cite{landau1977,vol10,simon1976}.  When the attractive part is very
weak we can appoximate the wave function in eq.~(\ref{funct-small}) by
the first term. This 
gives
\begin{equation}
 \lambda_+^{cr}=-\lambda_-\frac{\int\mathrm{d}r r V_-(r)}
{\int\mathrm{d}r r V_+(r)}+O(e^{-\frac{c}{\lambda_-^2}}).
 \label{weak-pot}
\end{equation}
Then from eq.~(\ref{condition}) we see that when
$\lambda_+\leq\lambda_+^{cr}$ the net volume 
 is less than $O(e^{-\frac{c}{\lambda_-^2}}) \ll  \lambda_-$ and
at least one bound state is present, whereas the system is unbound for
$\lambda_+>\lambda_+^{cr}$ where the net volume is positive.
This result leads us to a second necessary condition for a Borromean system,
that is $g\int V r \mathrm{d}r>0$, as only those potentials 
have a region of $g$ without two-body states.Second, 
we consider the two delta-shell potentials, $gV=\hbar^2/(2\mu d^2)\big(\lambda_+
\delta(r/c-1) -\lambda_-\delta(r/d-1))$, which is infinitely large for a
given radial distance and zero otherwise.  In the limit when
$c\rightarrow d$ we immediately conclude from eq.~\eqref{condition}
and continuity of the wave function that $\lambda_+^{cr}= -
\lambda_-$.

Other potentials localize the wave
function in the attractive region, which increases
$\int\mathrm{d}r rV_-(r)\phi(0,r)$ and  decreases $\int\mathrm{d}r r V_+(r)\phi(0,r)$.
Eq.~(\ref{condition}) then strongly suggests a steep
increase of $\frac{\mathrm{d}\lambda_+^{cr}}{\mathrm{d}\lambda_-}$ at
a given sufficiently large value of $\lambda_-$. Or, in another words,
if the potential is of finite range then for 
$\lambda_+^{cr}\rightarrow \infty$ a
finite $\lambda_-^{cr}$ exists that will bind the two-body system.

\subsection {Square well with barrier or core}

The threshold conditions are not easily derived for arbitrary
potentials with sizable attraction and repulsion.  However, the
general equations, eqs.~\eqref{condition} and \eqref{funct-small}, are
directly applicable for simple potentials where zero-energy wave
functions are known.  From properly selected potentials we can then
extract correct qualitative properties for more general potentials.  We choose to study a solvable model
containing all the crucial features, that is
\begin{eqnarray}
V=\lambda_-V_-+\lambda_+V_+ =\left\{ \begin{array}{ll}
  - \frac{\lambda_-\hbar^2}{2\mu R_s^2} &  r\leq R_s   \\
   \frac{\lambda_+\hbar^2}{2\mu R_s^2} &  R_s< r \leq R_l \\
   0        &   r>R_l  \;,
\end{array}\right.
\label{sqpot}
\end{eqnarray}
where $R_s$ and $R_l$ are positive radii. 
This potential is useful 
for describing qualitative features of more general potentials, see e.g. ~\cite{jeremy2010}.
 The zero-energy
solution then has the form
\begin{equation}
\phi(0,r)=\left\{ \begin{array}{ll}
AJ_0(k_1r) & \; r\leq R_s \\
BI_0(k_2r)+CK_0(k_2r) & \;R_s< r \leq R_l  \\
DK_0(k_3r) &  \; r > R_l   \;,
\end{array}\right.
\label{WFsa}
\end{equation}
where $k_1=\sqrt\lambda_-/R_s$, $k_2=\sqrt\lambda_+ /R_s$, $k_3 \rightarrow 0$,
$J_0,I_0,K_0$ are Bessel functions with the usual definitions
\cite{abram64}, and the constants $B,C,D$ are defined through $A$ by
matching at the points $r=R_s$ and $R_l$.  One of these equations is
the quantization condition providing the energy $E$.  For $E=0$ we use
instead eq.~\eqref{condition}, which by use of
$xZ_0(x)=\frac{\mathrm{d}}{\mathrm{d} x} x Z_1(x), Z_i=I_i,Y_i,J_i$,
$xK_0(x)=-\frac{\mathrm{d}}{\mathrm{d} x} x K_1(x)$ and the wave
function in eq.~\eqref{WFsa} can be integrated to give
\begin{eqnarray}
aI_1(k_2R_l)=bK_1(k_2R_l) ,\;
 \label{box-pot}
\end{eqnarray}
where the constants $a$ and $b$ are defined by
\begin{eqnarray}
a&=&k_2  J_0 (k_1 R_s)K_1(k_2 R_s)-k_1  J_1(k_1R_s)K_0(k_2R_s), \nonumber \\
b&=&k_2  J_0 (k_1 R_s)I_1(k_2 R_s)+k_1 J_1(k_1R_s)I_0(k_2R_s).
\label{const}
\end{eqnarray}
The derivation of eqs.~\eqref{box-pot} and \eqref{const} employed the
bound state boundary condition and the relations
$J_0(x)Y_1(x)-J_1(x)Y_0(x)=-\frac{2}{\pi x}$ and
$I_0(x)K_1(x)+I_1(x)K_0(x)=\frac{1}{x}$.

First we start with overall weak potentials, that is
$k_2R_s \rightarrow 0$ and $k_1R_s \rightarrow 0$.  Then the
properties of the Bessel functions 
give $a\rightarrow 1/R_s$ and
$b\rightarrow R_s(k_1^2+k_2^2)/2$.  Using that 
$J_1(x)\approx x/2, K_1(x)\approx 1/x$ we
find that eq.~\eqref{box-pot} is equivalent to $\int V r \mathrm{d}r =
0$, and coinciding with eq.~\eqref{weak-pot}.

The other limit of a sizable barrier where $k_2R_s\gg1$ can also be
found analytically.  After some manipulations using 
properties of the Bessel functions, we conclude that
Eq.~\eqref{box-pot} can be approximated by
\begin{eqnarray}
k_2R_sJ_0(k_1R_s)-k_1R_sJ_1(k_1R_s)= 0\;,
 \label{big-str}
\end{eqnarray}
where we assumed that $R_l\neq R_s$.  Thus,
the solution has to be near the nodes of the Bessel function, since
$J_0(k_1 R_s) =k_1R_sJ_1(k_1R_s)/(k_2R_s)\rightarrow 0$.  Continuity
of the wave function at $r=R_s$ combined with the exponential decrease
by penetrating into a barrier then in turn implies that the wave
function must be severely diminishing with increasing barrier.

\begin{figure}
\epsfig{file=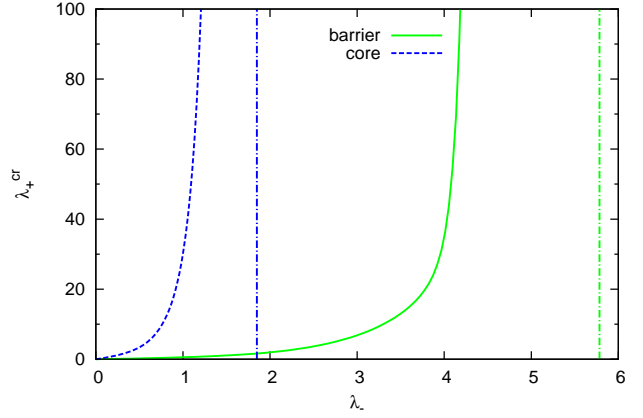,scale=0.95}
\caption{The critical strength, $\lambda_+^{cr}$, on the positive part
  of the potential as function of the strength, $\lambda_-$, on the
  negative part. We use $R_s/R_l=1/2$.  The two curves
  are for the square well with a barrier (solid green), eq.~\eqref{box-pot},
  and the inverted with a core and no outer barrier (dashed blue),
  eq.~\eqref{box-pot-flip}.  The vertical dot-dashed lines are the corresponding
  asymptotic values, $\lambda_-^{cr}$.  }
\label{fig2}
\end{figure}

The critical value, $\lambda_+^{cr}$, are shown in Fig.~\ref{fig2} as
function of $\lambda_-$, see eq.~\eqref{box-pot}.  The behavior is
typical and thus characterizes also the behavior for less schematic
potentials.  The linear behavior for weak potentials (small
$\lambda_-$) is as predicted in eq.~\eqref{weak-pot}. As $\lambda_-$
increases the critical repulsion rises steeply and at some point,
$\lambda_-^{cr}$, it cannot compensate to avoid a bound state. Larger
attraction, $\lambda_- >\lambda_-^{cr}$, always provides two-body
binding.

The potential in eq.~\eqref{sqpot} could as well be understood with
positive and negative parts interchanged. 
\begin{eqnarray}
V=\lambda_-V_-+\lambda_+V_+ =\left\{ \begin{array}{ll}
   \frac{\lambda_+\hbar^2}{2\mu R_s^2} &  r\leq R_s   \\
  -\frac{\lambda_-\hbar^2}{2\mu R_s^2} &  R_s< r \leq R_l \\
   0        &   r>R_l  \;,
\end{array}\right.
\label{sqpot1}
\end{eqnarray}
Then the corresponding
condition in eq.~\eqref{box-pot} has to be modified as achieved most easily 
by analytic continuation.  This implies the
use of $k_1=\sqrt{\lambda_+}/R_s,k_2=\sqrt{\lambda_-}/R_s$, resulting in
\begin{eqnarray}
\tilde a J_1(k_2 R_l) = \tilde b Y_1 (k_2 R_l)\;,
 \label{box-pot-flip}
\end{eqnarray}
where the constants correspondingly have to be changed to
\begin{eqnarray}
\tilde a&=&k_2 I_0 (k_1 R_s)Y_1(k_2 R_s)+k_1I_1(k_1R_s)Y_0(k_2R_s), \nonumber \\
\tilde b&=&k_2 I_0 (k_1 R_s)J_1(k_2 R_s)+k_1I_1(k_1R_s)J_0(k_2R_s).
\end{eqnarray}
For weak potentials we again find $\tilde a\rightarrow 1/R_s,
\tilde b\rightarrow (k_2^2R_s+k_1^2R_s)/2$, which as before leads to
the condition for binding with zero net volume.

\begin{figure}
\epsfig{file=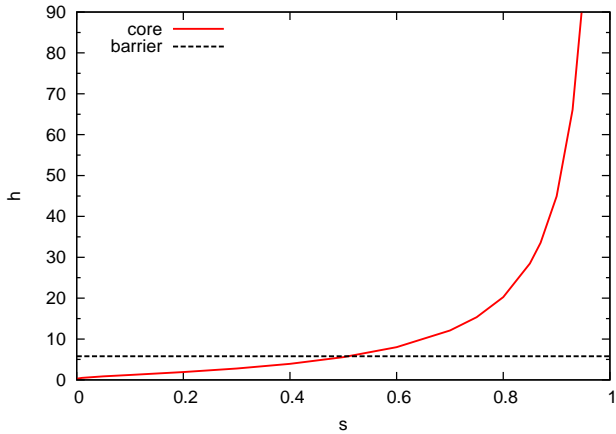,scale=0.95}
\caption{The dimensionless volume, $h$, of the attractive part as function of the
  ratio, $s=R_s/R_l$, of inner and outer radius for the square well potential
  with a repulsive inner core, see eq.~\eqref{barr}. The horizontal
  dashed line is the volume needed to bind two particles when an outer
  barrier is present, see eq.~\eqref{big-str}. }
\label{fig1}
\end{figure}

However, these potentials of finite size with $k_1R_s \gg 1$ behave
qualitatively different from the potential in eq.~\eqref{sqpot}.
In this limit
it is again possible to reduce eq.~\eqref{box-pot-flip} to give
\begin{eqnarray} J_0
  (k_2 R_s)Y_1(k_2 R_l)-J_1(k_2R_l)Y_0(k_2R_s)=0.
\label{barr}
\end{eqnarray}
This equation provides the boundary condition for a wave function
which is zero at $r=R_s$ and has zero derivative at $r=R_l$.
This condition for binding by an attractive square well then depends
on both $k_2R_s$ and $k_2R_l$. These two parameters can be expressed
in terms of the dimensionless measure of the attractive volume,
i.e. $h = k_2^2(R_l^2-R_s^2)$, and the ratio between the two
radii $s=k_2R_s/(k_2R_l)<1$, that is $k_2^2R_l^2 = h/(1-s^2)$ and
$k_2^2R_s^2 = h s^2/(1-s^2)$.  Expressed by $h$ and $s$,
eq.~\eqref{barr} determines $h$ as function of $s$ as shown in
Fig.~\ref{fig1}.  For $s\rightarrow 1$, $h \sim \frac{1}{1-s}$ 
and for small values of $s$: $h\sim-1/\ln s$. 
We note the continuous approach to the $s=0$ limit 
which corresponds to the binding in an overall attractive potential
where only infinitesimal strength or volume is necessary.  The other
limit of $s=1$ exhibits that the potential must be infinitely deep
when the two radii approach each other at a finite value.  This
variation is completely different from the condition of binding for
the potential from eq.~\eqref{sqpot} where the volume for deep
potentials is independent of the barrier dimension.
We also solve eq.~\eqref{box-pot-flip} for given ratio $R_l/R_s$
and exhibit $\lambda_+^{cr}$ as function of $\lambda_-$ in Fig.~\ref{fig2}.   
Again we find slow convergence towards the asymptotic value obtained from
eq.~\eqref{barr}.

In 1D we consider potentials given by eqs.~\eqref{sqpot} and \eqref{sqpot1} 
for $x>0$ and infinite wall for $x<0$.
We write then, instead of eq.~\eqref{big-str}, $k_1R_s=\pi$, where
only $k_1R_s$ is present and eq.\eqref{barr} is replaced by
$k_2(R_l-R_s) = \pi/2$, where only the difference $R_l-R_s$ appears
due to translational invariance \cite{landau1977}.  
In 3D we get a radial equation which is the same as in 1D.
Moreover we also have the same boundary conditions, because 
we assume that 1D potentials have an infinite wall for negative $x$. 
Thus, in 3D we get the threshold conditions,$k_1R_s=\pi$ and $k_2(R_l-R_s) = \pi/2$
for potentials from eqs.~\eqref{sqpot} and \eqref{sqpot1} respectively.

So far we can conclude that only potentials with positive volume integral
can provide Borromean states. Also we see that the strength
of the attractive part, $\lambda_-$, can not be larger than $\lambda_-^{cr}$,
as $\lambda_->\lambda_-^{cr}$ always produce a two-body bound state. 
This is the knowledge that we can extract without solving the three-body problem.

\section {Three particles in 2D}

We consider the three-body
Schr{\"o}dinger equation for three identical bosons
with $\lambda_{+}$ larger than $\lambda_{+}^{cr}$.
If this system has a bound three-body state 
we call it Borromean. 
To make this procedure formal we define $\Lambda_+^{cr}(\lambda_-)$ 
as the threshold value of $\lambda_+$ for binding of the three-body system
(for a given $\lambda_-$, potentials with $\lambda_+>\Lambda_+^{cr}$
can not provide three-body bound states).  
Since $\Lambda_+^{cr} \geq \lambda_+^{cr}$, Borromean states then appear if $\Lambda_+^{cr}\neq
\lambda_+^{cr}$ for repulsive strengths in the interval
$\lambda_+^{cr}<\lambda_+<\Lambda_+^{cr}$. We want to establish
which potentials can provide Borromean binding.

\subsection {Simple cases}

It is entirely possible that Borromean systems do not
exist for some shapes of the interaction, that is $\Lambda_+^{cr} = \lambda_+^{cr}$.  We shall here
provide a number of such examples, some of which we have already mentioned.  

First, {\it purely attractive potentials}, $gV<0$, where the two and
three-body thresholds coincide, $g\rightarrow 0$ \cite{tjon1980,nielsen1999}.
Second, {\it potentials with negative or zero net volume} but not necessarily weak.  These
potentials have at least one bound two-body state for all, even
infinitesimally small, strengths \cite{landau1977,vol10,simon1976}.
Thus, again the two and three-body thresholds are the same and
Borromean states cannot exist. For the same reasons these potentials
can not provide Borromean binding also in 1D. 
The third example is the {\it delta shell potential}, 
\begin{equation}
gV=\hbar^2/(2\mu d^2)\big(\lambda_+
\delta(r/c-1) -\lambda_-\delta(r/d-1)\big), 
\end{equation}
with $c\rightarrow d$.  Consider a potential
with three bound particles, i.e. $\lambda_+ < \Lambda_+^{cr}$.  
Ref.~\cite{richard1994} proves that if  
three particles are bound with a potential $gV$, then two particles are bound
with a potential $\frac{3}{2}gV$ , thus the potential
\begin{equation}
\hbar^2/(2\mu d^2)\big(\frac{3}{2}\lambda_+ \delta(r/c-1) -\frac{3}{2}\lambda_-\delta(r/d-1)\big),
\end{equation}
with $c\rightarrow d$ binds two particles.  From the discussion in
connection with eq.~\eqref{weak-pot} for two particles, we then know
that $\lambda_+ < \lambda_+^{cr} = \lambda_-$, and as $\lambda_+
\rightarrow \Lambda_+^{cr}$ we therefore find $\Lambda_+^{cr} =
\lambda_-$.  Again, Borromean systems do not exist for this potential.

Surprisingly, even potentials with positive net volume do not always 
provide Borromean binding. First it was pointed out in Ref.~\cite{cabral1979}
that the Lennard-Jones potential have dimer and trimer thresholds 
for the same coupling constant.  Moreover numerical search for 
three-body bound states with potentials of the form 
\begin{equation}
gV(r)=(\hbar^2/2\mu b^2)[-\alpha_1 e^{-r^2/b^2}+\alpha_2 e^{-4r^2/b^2}],
\end{equation}
with $\alpha_i>0$
shows absence of Borromean binding \cite{tjon1980,nielsen1999} . 
This has to be compared with the 3D behaviour, where those potentials 
always have a regime, when three but not two particles are bound \cite{richard2000,cramer1977}.
Further extensive numerical investigation 
carried out in Ref.~\cite{nielsen1999} yielded an example of potential
$gV(r)=(\hbar^2/2\mu b^2)[2 e^{-r^2/(2 b^2)}-5.7 e^{-2r^2/b^2}]$ that can provide 
Borromean binding. 
Another example with Borromean binding was suggested for 3D in \cite{richard2000}
but it is also useful in 1D and 2D 
\begin{equation}
gV(r)=\left\{ \begin{array}{ll}
g\frac{\hbar^2}{2\mu b^2}(\frac{r^2}{2b^2}-1) & \; r \leq C\times b \sqrt{\frac{1}{g}} \\
f(r) & \; r>C\times b \sqrt{\frac{1}{g}} \; ,
\end{array}\right.
\label{osc_pot}
\end{equation}
where $f(r)$ is a positive continuous function that vanishes at infinity.
We see that if the constant $C$ is sufficiently large we obtain 
Borromean binding for $\frac{4}{3}=g_3^{cr}<g<g_2^{cr}=2$. This potential
can be used in any dimension to obtain the largest possible Borromean window ($g_3^{cr}/g_2^{cr}$).
 Moreover eq.~\eqref{osc_pot} suggests that Borromean systems can be obtained 
for any decay of the potential at infinity.
For example, potentials with $1/r$ behaviour at infinity, that we 
do not consider above, and deep enough pocket can produce Borromean binding.

The overall conclusion is that to get
Borromean binding we need potentials with positive net volume which leads
to finite strength, in contrast to (infinitesimally) small. Moreover, 
the examples we provided above show that so far we do not know potentials
without an outer barrier that can produce Borromean binding. 
It might mean that only potentials with an outer barrier 
are able to provide Borromean states in 2D.

\subsection {Three-body conditions}
\label{three_body}

To get Borromean binding  we need the three-body wave function 
to be localised in the region where all three
particles are close to each other. However localization
increases the kinetic energy. This interplay between 
kinetic and potential energies defines the possibility
for existence of Borromean states.
We illustrate it by using  the hyperspherical expansion method,
which is efficient for weakly bound systems \cite{fedorov1993,nielsen1997}.  
An upper bound for the energy is found by use of just 
the lowest hyperradial potential
\begin{equation} \label{hyprad}
\bigg(-\frac{\partial }{\partial \rho^2}+\frac{3/4}{\rho^2}+
\frac{2m}{\hbar^2}(V_{\mathrm{eff}}(\rho) - E)\bigg)f(\rho)=0 \;,
 \end{equation}
where $E$ is the energy, $m$ the mass of the particles labeled
$(1,2,3)$, $V_{\mathrm{eff}}$ is the hyperradial potential, and the hyperradius
is an average length coordinate defined by $3\rho^2 = \sum_{i<k} {\bf
  r}_{ik} ^2$, where ${\bf r}_{ik} = {\bf r}_i - {\bf r}_k $. We also
have $\rho^2 = \frac{1}{2} r_{12}^2 + \frac{2}{3} r_{12,3}^2 $ and
$r_{12,3}^2 = \frac{3}{4} (r_{13}^2 + r_{23}^2 )$.  
The eq.~\eqref{hyprad} is well-known for $3D$ \cite{fedorov1993}, 
where $3/4$ is replaced by $15/4$.

First, we consider weak purely attractive potentials in $2D$ that support
only one two-body bound state with energy $E_2$ and root mean square (rms) radius 
\begin{equation}
R_2=\sqrt{-\hbar^2/(3\mu E_2)}.
\end{equation}
In this case, the effective potential, $V_{\mathrm{eff}}$, 
supports only two three-body bound states \cite{bruch1979,nielsen1997,kart2006,hammer2004,hammer2011}
with energies $E_3=16.52 E_2$ and $E_3=1.27 E_2$ and rms radii $R_3=0.305 R_2$ and $R_3=2.55 R_2$. 
Those states are independent of the details of the interparticle interaction
for sufficiently small $E_2$ and are determined by the long-range behaviour of $V_{\mathrm{eff}}$,
which is defined by $E_2$. It follows that
these two universal states always exist if $\lambda_+ \rightarrow \lambda_+^{cr}$.
Consequently Borromean binding occurs when we have a third state  
with rms radius proportional to the range of the interaction.
Then two bound three-body states move into the continuum (become unbound) with 
$E_2\rightarrow 0$, but the ground state at smaller distance remains bound.  
This mechanism for Borromean states reflects that the short-range part
of the potential $V_{\mathrm{eff}}$ is responsible.

\begin{figure}
\epsfig{file=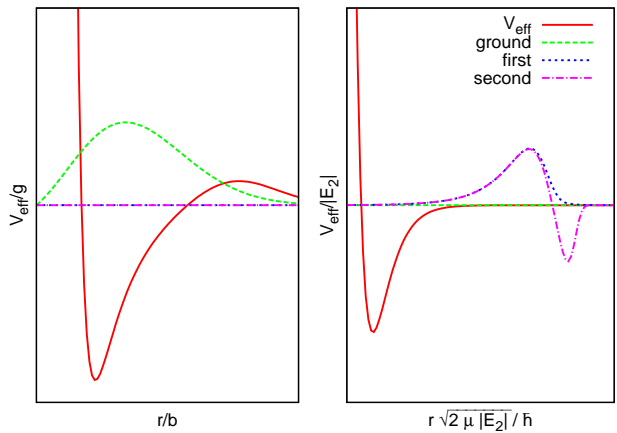,scale=0.95}
\caption{Division of the lowest 
effective adiabatic potential $V_{\mathrm{eff}}$ (solid line) for $E_2 \rightarrow 0$ into 
$V_{\mathrm{sh}}$ (left) and $V_{\mathrm{lon}}$ (right).
Range and strength of $V_{\mathrm{sh}}$ are defined 
through the range $b$ and strength $g$ of the two-body potential. 
Range and strength of $V_{\mathrm{lon}}$ are universal   
and defined through the two-body energy $E_2$. 
$V_{\mathrm{lon}}$ supports two universal three body states (dotted and dash-dotted lines).
Here $V_{\mathrm{sh}}$ supports a third bound state (dashed) which becomes Borromean for $E_2 = 0$.
If $V_{\mathrm{sh}}$ cannot support a bound state for $E_2 = 0$, a Borromean state does not occur.}
\label{fig4}
\end{figure}

Let us now make an explicit division of $V_{eff}$ into two parts
$\frac{2m}{\hbar^2}V_{\mathrm{eff}}(\rho)=V_{\mathrm{sh}}(\rho)+V_{\mathrm{lon}}(\rho)$, where $V_{\mathrm{lon}}$ depends just on two-body 
binding energy and may support the two weakly
bound three-body states, and $V_{\mathrm{sh}}(\rho)$ depends on the details of the interparticle 
interaction. We illustrate this division on 
Fig.~\ref{fig4}. A crude way to estimate the interplay between kinetic energy
and potential energy is by using
the short-range part $V_{\mathrm{sh}}$ in eq.~\eqref{hyprad} instead of $V_{\mathrm{eff}}$.
Appearence of the bound state in such an equation is discussed in Ref.~\cite{gibson1986}, that is
\begin{equation}
1+\int_0^\infty \mathrm{d}\rho \frac{\phi_1(0,\rho)V_{\mathrm{sh}}(\rho)}{\sqrt{\rho} } =0,
 \label{condition-3-particles}
\end{equation}
where the function $\phi_1(0,\rho)$ is defined through 
\begin{eqnarray}
\phi_1(0,\rho)=\frac{\rho^{3/2}}{2}\bigg(1+\int_0^{\rho}\mathrm{d}\rho' 
\frac {\phi_1(0,\rho') 
V_{\mathrm{sh}}(\rho')}{\sqrt{\rho'}}\bigg) \nonumber \\
- \frac{1}{2\sqrt{\rho}}\int_0^{\rho}\mathrm{d}\rho'\rho'^{3/2}\phi_1(0,\rho')
V_{\mathrm{sh}}(\rho').
 \end{eqnarray}

Eq.~\ref{condition-3-particles} directly expresses that a vanishing
strength of $V_{\mathrm{sh}}$ cannot satisfy this equation. Thus, since a
Borromean system requires a bound state, the potential must have a
finite strength.  It means, for example,
that square wells from eq.~\eqref{sqpot1} with
$R_s/R_l\rightarrow 0$ can not provide Borromean binding because 
they need a very small depth of the pocket to provide a two-body bound state. 

\subsection {Square well potentials}

Eq.~\eqref{condition-3-particles} qualitatively expresses 
the idea of the interplay between kinetic energy and interaction.
A deep enough attraction inside a repulsive barrier, e.g. 
eq.~\eqref{sqpot}, allows Borromean systems,
since all three particles can benefit simultaneously from the 
attraction, without extending into the barrier. This is discussed for
a schematic case in the Appendix.
One would expect that potentials without an outside barrier, e.g. eq.~\eqref{sqpot1},
also must support Borromean binding for some $\lambda_-,\lambda_+^{cr}, R_s/R_l$.
However extensive numerical searches did not reveal Borromean binding. 
Unfortunately, exploiting numerical search, we can not exclude it rigorously 
for all potentials without an outer barrier. 
However, using a rough
estimate obtained from eq.~\eqref{condition-3-particles}  we 
can suggest a region where Borromean 
system might occur. To do so we establish 
a qualitative connection between a two-body square-well 
potential with infinite core
and an effective three-body potential.  We want to use only
simple solvable potentials to suggest
possibilities that unravel the general trend (see the appendix).

\begin{figure}
\epsfig{file=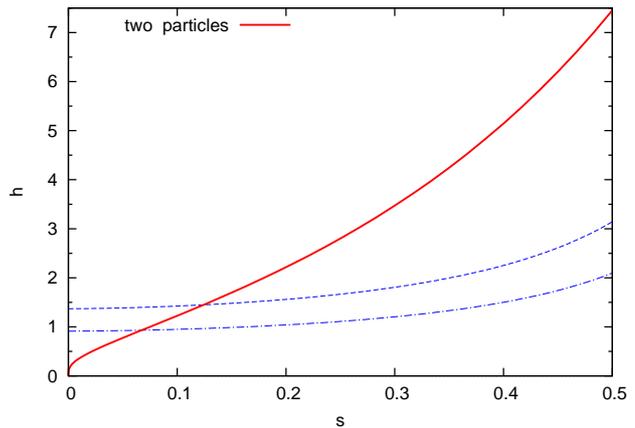,scale=0.95}
\caption{Axis and the solid red line is the same as on Fig.~\ref{fig1}.
Dashed and dot-dashed blue lines give an insight into the regions where three-body 
bound states with root mean square radius proportional to the range of the potential 
(see the Appendix) might exist.
The space between solid and dashed (dot-dashed) line (when solid line is higher)
gives a rough estimate for $h$ and $s$ that can support Borromean states.}
\label{fig2a}
\end{figure}

Using this connection we determine parameters of the 
two-body potentials that give three-body bound states
with rms radius proportional to the range of 
the potential. To determine whether this is a Borromean state, 
we compare parameters of this potential with a two-body potential, that
satisfies eq.~\eqref{barr}.  
We present the result of this qualitative investigation on 
Fig.~\ref{fig2a}, where two-body bound states
exist above the solid line, while dashed and dot-dashed lines 
represent parameters of the two-body potential that might allow
three-body ground state with rms radius 
proportional to the range of the potential (see the Appendix). Thus 
Borromean systems are likely to exist below the two-body and above the three-body
curves. The optimum configuration occurs when the two-body
attraction is most effective for all pairs. By simple geometric 
considerations, this dictates that
$R_l$ must be larger that $2 R_s$. On the other hand $R_s/R_l$
can not go to zero, where as follows 
from Fig.~\ref{fig2a} a two-body system is easily bound. This analysis 
suggests that the region where $R_l \sim 2 R_s$ has to produce the largest  
Borromean window, because the negative part of the potential
is deep enough to overcome the kinetic energy term, and radii do not
deviate too much from each other, which allows the most effective attraction. 
This qualitative procedure unravels regions where 
three body Borromean state 
are most likely to exist. 

\subsection {Numerical illustrations}

\begin{figure}
\epsfig{file=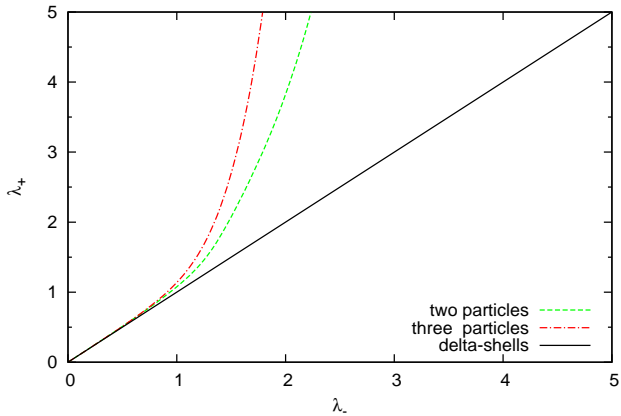,scale=0.95}
\caption{The two- and three-body critical strengths, $\lambda_+^{cr}$ (dashed line)
  and $\Lambda_+^{cr}$ (dot-dashed line), on the positive part of the potential as
  function of the strength on the negative part.  The potential is
  $gV=\lambda_+V_++\lambda_-V_-$ where $V_{\pm}$
  are positive and negative parts of 
 $\frac{\hbar^2}{2\mu b^2}\big(\exp(-0.5r^2/b^2) - 2 \exp(-2r^2/b^2)\big)$, respectively. 
  Solid line represents results for delta-shell potential, i.e. $\lambda_+^{cr}=\lambda_-$.}
\label{fig3}
\end{figure}

Our numerical procedure is based on the stochastic variational method
with basis of correlated Gaussians. This procedure was proven to give
accurate results \cite{suzuki1998}, however the convergence slows down
for potentials with a hard core. 
Numerically convenient potentials
are linear combinations of gaussians, because it allows calculation of matrix 
elements analytically. Moreover extensive numerical search with 
different potentials did not reveal Borromean states without an outer 
barrier. Thus, we choose a potential with outer barrier in the form
$V_{\pm}=\frac{F\pm |F|}{2}$, with 
$F=\frac{\hbar^2}{2\mu b^2}\big(\exp(-0.5r^2/b^2) - 2 \exp(-2r^2/b^2)\big)$,
where the net volume is zero. This potential is attractive for $r^2/b^2<\frac{2}{3} \ln 2 $ 
and repulsive otherwise.
  We show in Fig.~\ref{fig3} the critical
values for binding the two- and three-body systems as function of the
attractive strength, $\lambda_-$. The straight line for the
delta-shell potential is followed for small $\lambda_-$ for both two
and three particles.  The zero net volume behavior of $\lambda_+^{cr}
\approx \lambda_-$ is found as expected in this region.
All three curves begin to deviate around $\lambda_- \approx 0.7$, and
the finite range potentials reveal their divergent character.  We
define the critical values, $\lambda_-^{cr}$ and $\Lambda_-^{cr}$,
above which two and three-body bound states respectively are present
independent of $\lambda_+$. This limit can be described by potentials
of the form eq.(\ref{osc_pot}) and we see numerically the
expected ratio, $\Lambda_-^{cr}/\lambda_-^{cr} \sim 2/3$.  When $\lambda_-
> \lambda_-^{cr}$ the attractive pocket alone supports a two-body bound state.

At the
threshold $\lambda_+ = \lambda_+^{cr}$ two three-body states disappear
into the continuum. They have the same threshold as the two-body
system in complete analogy to weakly bound systems where the two and
three-body thresholds are identical \cite{nielsen1999,bruch1979}.
The difference for the deep potentials is that the Borromean state
remains.

\section{Experimental realization}
From the discussion above it is obvious that Borromean systems 
in 2D are not as abundant as in 3D. In particular, we need
an outer barrier and an attractive pocket. Systems where there is no
outer barrier and those with non-positive net volume integral
of the interaction are therefore out of the picture. This 
includes the case of neutral atoms in a plane
which would have van der Waals-type attractive pockets and 
inner hard-core repulsion. A popular way to achieve long-range
interactions is to use polar molecules \cite{lahaye2009,baranov2012}.
In particular, layered systems with polar molcules have promising 
properties and have recently been stabilized experimentally \cite{miranda2011}. 
The geometric dipole-dipole interaction of polar 
molecules in a multilayer system supports a number of 
interesting few-body 
states \cite{jeremy2010,wang2006,klawunn2010,zinner2012-1,jeremy2012,artem2012}
that are indicators of non-trivial many-body 
pairing \cite{potter2010,pikovski2010,zinner2012-2}. However, 
interlayer interactions always have zero net volume integral
\cite{jeremy2010}, while the intralayer interactions are 
purely repulsive \cite{jeremy2012} and do not
have the necessary pocket plus barrier structure.

In the following we list possible 
experimental setups where it could be possible 
to simultaneously maintain an outer barrier and
a substantial inner attractive pocket.
 
The first option is to use cold ionized atoms 
in a two-dimensional geometry. Ideally this would be ions confined to live on a 
surface. Other interesting setups could be ions trapped near a 
surface \cite{szymanski2012} or in a Penning trap \cite{britton2012}.
While these studies are mostly done with atoms in a crystal state, 
the few-body dynamics studied in the current paper requires 
that the ions also have motional degrees of freedom that are 
continuous in some range (or at least quasi-continuous if 
there is some weak confinement in the 2D plane of motion).
To reach the Borromean regime, we have the outer barrier
from the Coulomb repulsion of the ions. The inner attractive
pocket would then need to be provided by short-range van der
Waals interactions. This pocket is then required to sit 
at some small yet not too small distance in order to 
avoid the regime where everything is controlled by chemical
reaction dynamics. What is particularly complicated about 
this proposal is the fact that the outer repulsive 
Coulomb barrier could be extremely large and render any
inner attractive pocket irrelevant. This could possibly 
be counteracted through electron screening that 
lowers the Coulomb barrier. Here we 
imagine that a plasma of ions and electrons could be 
useful if it can be confined to 2D. A more realistic 
possiblity could be mobile impurities in solid-state 
system where the background electron density provides
a screening effect.

A second, and presumably much easier, option is to use 
polar molecules in external fields. It has been predicted 
that applications of AC and DC fields 
in systems with particles that have non-zero permanent 
electric or magnetic dipole moments provides a way to 
taylor the inter-molecular 
interactions \cite{pfau2002,micheli2007,gorshkov2008,cooper2009}.
In a squeezed geometry (quasi-2D), AC fields can be used to 
control the existance of two-body bound states for 
both bosonic and fermionic molecules, and correspondingly 
may result in bound three-body states of AC field dressed 
molecules \cite{huang2012}. What is needed here is 
the presence of both outer barrier and inner attractive pocket. 
As is discussed in Ref.~\cite{micheli2007} this can be achieved
by applying both an AC and a DC external field to the system.
DC fields have been used to align polar molecules and 
control the overall magnitude of the dipole moment in 
two-dimensional systems \cite{miranda2011} and very recently
the application of AC fields to the same geometry has 
been reported \cite{neyenhuis2012}. A combination of these
two external influences could provide the potential 
profile necessary to produce and observe Borromean 
bound states in a 2D setup.

\section {Conclusions}
Two-body binding is easier achieved in 2D compared to 3D, because of the negative
centrifugal barrier in the radial two-body
Schr{\"o}dinger equation. 
In all other spatial dimensions the
centrifugal barrier is zero or positive.  This feature makes
2D geometry a case of special interest. 

We investigate possible few-body structures in 2D with
the goal to find conditions for Borromean states. 
Thus, we search for the potentials that can support three, but not two-body bound states.    
Our focus is on potentials with positive and negative parts, 
as purely attractive potentials in 2D always 
support two-body bound state thus excluding Borromean states.

We show that the necessary condition for a potential to provide Borromean states
is a substantial repulsive and attractive part.
This is a consequence of the fact that near a two-body threshold a Borromean state
is localised at small radii. This can happen only 
for potentials with substantial attraction to outweigh the
kinetic energy of localisation. 
Moreover, our numerical search indicates that 
potentials without outer barrier are highly unlikely 
to support three-body states without a two-body state.
 
We conclude that experimental observation of Borromean 
systems in 2D is possible only for interactions that have
an outer barrier, a substantial attractive region, and positive net volume.
Polar molecules or ions 
in squeezed geometries are potential candidate 
systems for the occurence of low-dimensional Borromean states. A 
more speculative possibility is to look for Borromean signatures 
among impurities on a solid-state surface. The details of the 
experimental scenarios and which options are more viable goes 
beyond the current principle discussion and will be 
the focus of future studies.

{\bf Acknowledgements}
We thank F.~F. Bellotti who was the first reader of the manuscript. This 
research was supported by grants from the Danish Council for Independent Research 
| Natural Sciences.
 
\appendix

\section{Connections to the Hyperspherical Expansion}

Here we qualitatively discuss the connection between an interparticle interaction and 
and the lowest adiabatic potential in the hyperspherical expansion method.

The hyperradial potential for three bosons interacting pairwise
through a square well of radius $R_0$ and depth $V_0$ is determined
semi-analytically for $s$-waves by solving trancendental algebraic
equations \cite{jen96}.  The hyperradius, $\rho$, is defined
through two-body distances between pairs of the three particles which
implies that the $3D$-procedure is directly applicable for $2D$ as
well. The method divides the $\rho$-coordinate into four intervals, that
is

(i) with $\rho < R_0/\sqrt{2}$, where all three two-particle distances
are smaller than $R_0$. 

(ii) $ R_0/\sqrt{2} < \rho < R_0\sqrt{2/3}$, where at least 2 pairs
interact, but configurations of 3 interacting pairs are also possible.

(iii) $R_0\sqrt{2/3} < \rho < R_0\sqrt{2}$, where at most 1 pair
interacts, but configurations of 3 non-interacting pairs are also
possible.

(iv) $ R_0\sqrt{2} < \rho $, where at least 2 pairs don't interact.

The hyperradial attractive potential in turn increases from $3V_0$ for
$\rho=0$ through various continuous steps to a value less than $V_0$
at $\rho= R_0\sqrt{2}$. For larger $\rho$-values the potential
approaches zero as $1/\rho^2$ until the scattering length is reached for
$\rho \approx a \gg R_0$.  For $\rho$ larger than $a$ the potential
approaches zero even faster, that is in $3D$ as $\propto a/\rho^3 $
and in $2D$ as $\propto a/(\rho^3\ln(\rho) ) $.  These potentials
must be supplemented by the centrifugal barrier term, which in $3D$
and $2D$ corresponds to effective angular momenta of $l^*=3/2$ and
$l^*=1/2$, respectively.

The large-distance behavior in $3D$ allows a number of bound states
provided the scattering length is sufficiently large. This means that
it is possible to vary $R_0$ and $V_0$, where the scattering length is
maintained to be much larger than $R_0$ while the volume is too small to
support  a bound two-body state.  By increasing $a$ towards infinity,
still for an unbound two-body system, the number of three-body bound
states increase logarithmically towards infinity.  This is the Efimov
effect.

In $2D$ any, even infinitesimal, overall attractive potential provides
a bound state. Thus, when the three-body potential can bind with the
$l^*=1/2$-term, also the two-body system with $l^*=-1/2$-term is
bound. No Borromean system can be constructed, and no Efimov effect
exists, see subsection~\ref{three_body}.

We consider now the more general two-body potential of an attractive
square well of radius $R_s$ and depth $V_s$, and a repulsive barrier
between $R_s$ and $R_l$ of height $V_l$. The most interesting
combination in the present context is when $R_l > 2R_s$, since then
the large-distance configurations apply for the $s$-part of the
potential before the short-distance behavior of the $l$-part has
ceased to be present.  In other words, the attraction contributes fully
for all pairs when $\rho<R_s/\sqrt{2}$.

The volume, $h_3 = k^2 (R_l^2-R_s^2)$, for the three-body potential is
in general much larger than, $h_2$, for the two-body potential. We can
roughly relate by $h_3 \approx h_2 f_{\mu} f_{V} f_{R} $, where
$f_{\mu} = m/\mu =2$ is the ratio of the mass, $m$, in
eq.(\ref{hyprad}), $f_{V} = \langle V\rangle/V_0 \approx 2$ is the average
three-body potential in units of $V_0$, and $f_{R} \approx
(R_0\sqrt{2}/R_0)^2 = 2$. Increase of the volume leads to the possibility
of the Borromean binding. 

In the limit of a huge barrier, $V_l$ is very large, we can then
evaluate the condition for a bound three-body state.  The hyperradial
wave function, $J_{l=1}(\kappa \rho)$, for $l^*=3/2$ must have a node
at $\rho = R_s/\sqrt{2}$, where $\kappa$ is the corresponding wave
number. The dimensionless volume ratio is then, $h_3/h_2 = f_{\mu}
f_{V} f_{R} = 2 \times 3 \times 1/2 = 3$. The node, $14.66$, is
smaller than $3$ times the corresponding node, $ 5.78$ of the two-body
wave function, $J_{l=0}(kr)$, which immedately implies that there is a
window where three, but not two, particles are bound. Borromean
systems exist when $4.89<h_2<5.78$, that is for small $s<0.5$.

Changing the sign of the two-body potential gives a repulsive core up
to $R_s$ and afterwards until $R_l$ an attractive well. We use again
the relation between two and three-body potentials where three-body
binding must arise from the attractive part in the interval
$R_s\sqrt{2} < \rho < R_l\sqrt{2} $.  The limit of $s \rightarrow 0$
correspond to two-body potentials resembling
an overall very weakly attractive potential. 
In this case Borromean systems cannot exist, see subsection \ref{three_body}.

Assuming an average
potential strength of $2V_l$ we get, with $s_3 = s_2 = R_s/R_l$ that
Borromean systems exist when $8h_2>h_3$.  This comparison can be
improved by replacing $2V_l$ by a weighted average of the attractive
strength, $\langle V\rangle \approx \int V r dr / \int r dr =
19(1-36s^2/19)/[12(1-s^2)]$.  The average decreases with $s$ to a
value of $10/9$ for $s=1/2$.  This curve, shown in Fig.\ref{fig2a} - dot-dashed line,
allows a rather large window for Borromean states in an interval from
$0.1 <s < 0.5$.

These estimates of restrictions are rather conservative.  First they
are only estimates, and second two other, so far neglected effects,
oppose existence of Borromean states derived from the hyperspherical
formalism. The first is that the approximation to use only the lowest
adiabatic potential implies that the correct three-body energy must be
lower. This is derived from the general theorem that the correct
energy lie between the results from the lowest adiabatic potential
with and without the diagonal non-adiabatic term.  This tends to
further narrow down the Borromean window.

The second effect is that the fraction of the 2D
coordinate space where the distance between two particles is less than
$R_s$ is $R_s^2/(2\rho^2)$ for a given hyperradius.  Then one of the
attractions, $V_l$, should be replaced by the repulsion, $V_s$, in
total amounting to $V_l \rightarrow V_l (1- s^2(1+V_s/V_l)/3)$ when
$\rho = R_l/\sqrt{2}$.  For a larger $\rho = R_l\sqrt{2/3}$, the
factor $1/3$ should be replaced by $3/8$.  The requirement of an
overall positive volume of the two-body potential provide the
inequality $s^2(1+V_s/V_l) >1$, which gives a limit on the reduction
factor, that is at least $(1- s^2(1+V_s/V_l)/3) < 2/3$.  Thus, we
should further restrict the volumes by multiplying with $2/3$.  This
curve is also shown in Fig.\ref{fig2a} (dashed line), leaving still room for
Borromean states. However, a reduction by an additional factor of
$2$ would close that window completely.

In conclusion, the features of the square well must be maintained in
other potentials, that is with both attractive and repulsive
parts. The conclusions are therefore much more general.  Whether a
small window is open for Borromean states in $2D$ without a confining
outer barrier remains to be seen.  The features of candidate
potentials are that the ratio of attractive to repulsive volumes must
be between $8$ and $15$.  The ratio of repulsive to attractive
strengths must be limited to be smaller than $ 1/s^2-1 < V_s/V_l <
2/s^2-1$.


\begin{thebibliography}{99}
\bibitem{first_footnote} We consider potentials which are isotropic, fall off faster than $1/r^2$ at infinity, 
and are bounded. However, most of the results can be extended to potentials with infinite repulsion.
\end{thebibliography}
 \end{document}